\documentclass[a4paper]{article}
\usepackage{multicol}
\usepackage{amsmath}
\usepackage{mathrsfs}
\usepackage[utf8]{inputenc}
\usepackage{hyperref}
\usepackage{cite}
\usepackage{graphicx}
\usepackage[left=20mm, top=20mm, right=20mm, bottom=20mm, nohead, nofoot]{geometry}
\hypersetup{pdfstartview=FitH}

\newcommand{\s}[1]{\hat \sigma_{#1}}
\newcommand{\df}[1]{\text{d}#1}

\newcommand{\com}{\mspace{5mu}\text{,}}
\newcommand{\pnt}{\mspace{5mu}\text{.}}
\newcommand{\En}{\hat{\mathscr H}}

\usepackage[format=plain, font=footnotesize, singlelinecheck=false, labelfont=bf, labelsep=period]{caption}
\begin{document}
\onecolumn
\begin{center}
	\large{\textbf{On the plasmon dispersion in biased graphene bilayer}}
\end{center}
\vspace{1ex}
\begin{center}
	E.I. Kukhar $^\text {a, b }$\footnote{eikuhar@yandex.ru}, S.V. Kryuchkov$^\text {a, b }$
\end{center}
\vspace{1ex}
\begin{center}
	$^\text a$\textit{Volgograd State Socio-Pedagogical University, Physical Laboratory of Low-Dimensional Systems\footnote{ \url{http://edu.vspu.ru/physlablds}}, V.I. Lenin Avenue, 27, Volgograd 400066, Russia}
\vspace{2ex}
	
	$^\text b$\textit{Volgograd State Technical University, V.I. Lenin Avenue, 28, Volgograd 400005, Russia}
\end{center}

\begin{abstract}

Plasma oscillations in doped graphene bilayer at zero temperature has been investigated.
Bias voltage effect on the dispersion curve for plasmon in bigraphene has been studied in random phase approximation.
The possibility of controlling of curvature of dispersion curve for plasmon by changing of bias voltage has been shown.
The dependence of this curvature on the bias voltage has been predicted to have the nonmonotonous character.
Namely the existence of breaking point for such dependence has been found out.
The bias voltage corresponding to the breaking point is shown to increase with free carriers concentration as square root of concentration.
\\
\\
\\
\end{abstract}

\begin{multicols}{2}

\textbf{I. Introduction}\\

Modern achievements in the field of both solid state physics \cite{1} and electronics \cite{2} enable the manipulation of electronic properties of different condensed-matter structures with external electromagnetic field.
The aim of engineering based on such manipulation is creation of devises with tunable characteristics.
The essential attention is paid to the low-dimensional systems with Dirac-like dispersion.
These systems include graphene based materials.
However the gapless band structure of the graphene makes graphene-based materials inapplicable in the field of semiconductor electronics.

Presently bigraphene excites enormous interest in the physics of condensed-matter.
As compared with single graphene which electrons have the linear dispersion in the vicinity of Dirac point (K point) the electrons of bigraphene are the massive quasiparticles.
Besides there is a possibility of energy gap opening due to constant electric field applied perpendicularly to the graphene layers \cite{14}.
In this case electronic properties of gapped bigraphene essentially depend on the potential difference between the graphene layers (bias voltage).
As a consequence they are perspective for nanoelectronic applications.
Moreover the dispersion of electrons in biased bigraphene shows an unusual "sombrero-like" peculiarity \cite{14,15}.
Namely, energy minimum displaces from the K point and the region of the negative effective mass is formed in the vicinity of this point.
Both infrared \cite{16,17,18} and visible \cite{19,20} spectroscopy had been used to confirm these peculiarities of band structure of bigraphene.

The investigation of plasma oscillations and collective plasma excitations know as plasmons in graphene-based structures are of interest from both fundamental and practical point of view last time.
Fundamental interest is explained by the fact the plasmons are quantum-mechanical objects which have no analogouss in classical theory.
Moreover the account of interactions of charge carriers with elementary excitations in an number situations is important to explain the features of plasma oscillations in bigraphene \cite{32}, the formation of bound electron-plasmon \cite{29,30} and magnetoplasmon complexes \cite{31} in graphene.
The current progress in nanoplasmonics and optoelectronics explains the increased attention to the investigation of plasma waves in graphene structures \cite{33,34}.
The latter is due to the fact that the velocity of plasmons exceeds the electron drift velocities by several orders.
This should increase the speed of operation of nanoelectronic devices.

Plasma excitations in bigraphene had been investigated both within the hydrodynamic model \cite{35} and within random phase approximation (RPA) \cite{32,36,37,38}.
In Ref. \cite {39} the electron-phonon interaction had been taken into account.
Plasma oscillations in a doped unbiased bigraphene when the Fermi level was displaced from the K point had been studied in \cite{36,37,40} within RPA.
In Ref. \cite{38} the dispersion law for plasmons in the bigraphene with an intrinsic conductivity had been calculated in the case when the Fermi level had been located exactly at K point.
In this situation plasmons appeared only at nonzero temperatures.
The dispersion law of plasma oscillations in biased bigraphene (when inter-layer bias voltage $\varphi$ is nonzero) had been found within RPA in Ref. \cite{32}.
However in Ref. \cite{32} the electron gas had been assumed to be nondegenerate (the Fermi level lied inside the forbidden band).
\end{multicols}
\twocolumn
Besides the results of Ref. \cite{32} are valid for small bias 
($V_0\equiv e\varphi\ll \varepsilon_\perp$, $\varepsilon_\perp\simeq$390 meV is hopping integral between atoms which are on top of each other in neighboring layers)
because of the approximate expression had been used for the bigraphene energy spectrum.
Here we suggest the electron gas at zero temperature.
Also we suppose that Fermi level lies inside the allowed band of biased bigraphene.
In this case the electron gas of bigraphene can't be assumed as nondegenerate in contrast to Ref. \cite{32}. 

The paper is organized as follows.
In the section II we describe the energy spectrum of biased bigraphene.
In the section III we investigate the dependence of the plasmon dispersion on the bias voltage within the long-wavelength approximation.
To calculate the polarization operator we use the RPA and the energy spectrum for electron in bigraphene.
The last section is discussion.
\\
\\

\textbf{II. Energy spectrum for electron in biased bigraphene}\\

Let the bigraphene layers are parallel to the plane $xy$.
Quantum mechanical state of electron in bigraphene near one of the Dirac points is described by four-component spinor $\psi$ written in a basis corresponding to two crystal sublattices of graphene, and two graphene layers \cite{14}.
This spinor obeys the equation
$i\mspace{3mu}\partial_t\psi=\En\psi$
with the Hamiltonian matrix which within the tight-binding model can be represented in the form
\begin{multline}\label{2}
	\En=
	\frac{V_0}{2}
	\left(\begin{array}{cc}
		\hat 1 & 0 \\
		0 & -\hat 1
	\end{array}\right) +
	v_\text Fp_x
	\left(\begin{array}{cc}
		\s{x} & 0 \\
		0 & \s{x}
	\end{array}\right) \\
	+v_\text Fp_y
	\left(\begin{array}{cc}
		-\s{y} & 0 \\
		0 & \s{y}
	\end{array}\right) +
	\varepsilon_\perp
	\left(\begin{array}{cc}
		0 & \hat\eta \\
		\hat\eta & 0
	\end{array}\right) \com\mspace{30mu}
\end{multline}
where $v_\text F$ is velocity on the Fermi surface, $\hat 1$ is 2$\times$2 unit matrix,
$\hat{\boldsymbol\sigma}=\left(\s{x} \com\mspace{5mu} \s{y} \com\mspace{5mu} \s{z}\right)$
are Pauli matrices,
$$
\hat\eta=\left(\begin{array}{cc}
1 & 0 \\
0 & 0 \end{array}\right) \pnt
$$

\begin{figure}[t]
	\centering
	\includegraphics[width=\linewidth]{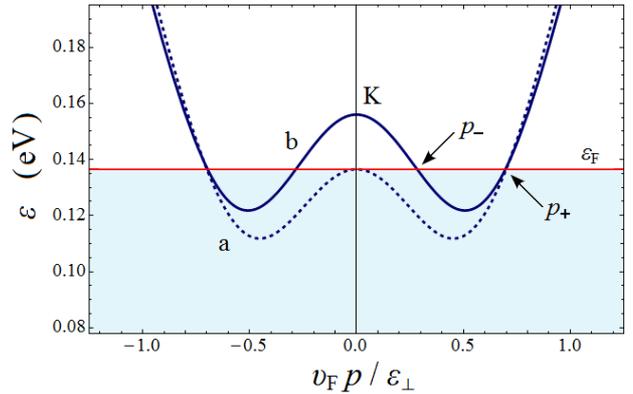}
	\caption{Shifting of the region of negative effective mass with increasing of the bias voltage;
		(a) -- $V_0=$312 meV; (b) -- $V_0=$273 meV}
	\label{fig1}
\end{figure}

Eigenvalue of the Hamiltonian \eqref{2} is \cite{14}
\begin{gather}\label{9}
	\varepsilon\left(\mathbf p\right)=\pm\sqrt{\frac{\varepsilon_\perp^2}{2}+\frac{V_0^2}{4}
		+v_\text F^2p^2\mp f\left(p\right)} \com
\end{gather}
$$
f\left(p\right)=\sqrt{\frac{\varepsilon_\perp^4}{4}+\left(\varepsilon_\perp^2+V_0^2\right)v_\text F^2p^2} \pnt
$$

The value of the gap in the energy spectrum described by Eq. \eqref{9} is
\begin{gather}\label{16}
	\Delta_\text g=\frac{\varepsilon_\perp V_0}{\sqrt{\varepsilon_\perp^2+V_0^2}} \com
\end{gather}
and the energy spectrum of electron in the vicinity of the K point is approximated by the expression (the lower branch of the conduction band)
\begin{gather}\label{17}
	\varepsilon\left(\mathbf p\right)\simeq\varepsilon_\text K
	-\dfrac{V_0v_\text F^2p^2}{\varepsilon_\perp^2}
	+\frac{v_\text F^4p^4}{\varepsilon_\perp^2V_0} \com
\end{gather}
where $\varepsilon_\text K=V_0/2$ is energy at K point.
The curvature of the dispersion line $\varepsilon\left(p\right)$ at K point and the value of the gap are seen from Eqs. \eqref{16} and \eqref{17} to increase with bias voltage.
\\
\\

\textbf{III. Dispersion law for plasmon in biased bigraphene}\\

In long-wavelength approximation the dispersion of plasmons in 2D structures is well known to have the form
$\omega_{\mathbf k}\propto\sqrt k$, where
$\mathbf k$ is wave vector of plasmon.
Proportionality constant determines the plasmon group velocity and depends on the electron properties of the material.
It is proportional to the curvature of the dispersion line
$\omega_{\mathbf k}$ at $k=0$.

Below the dispersion law for biased bigraphene is represented in the form
$\omega_{\boldsymbol\kappa}=\omega_0\sqrt{\kappa}$, where
$\boldsymbol\kappa=v_\text F\mathbf k/\varepsilon_\perp$
and $\omega_0$ is the function of bias voltage.
Calculation of the parameter $\omega_0$ is the main aim of this section.
Firstly, unlike Refs. \cite{36,37,38,40} we consider the biased bigraphene.
Secondly, in contrast to \cite{32} we use the more general electron spectrum which is also valid in the case when $V_0\sim\varepsilon_\perp$.
Thirdly, in opposite to Refs. \cite{32,38} the Fermi level is considered to lie inside the conduction band.

Within the RPA the plasmon dispersion is determined from the equation
\begin{gather}\label{10}
	1-V\left(\mathbf k\right)\Pi\left(\omega,\mathbf k\right)=0 \pnt
\end{gather}
The unscreened interaction is supposed to be the Coulomb interaction, so $V\left(\mathbf k\right)=2\pi e^2/k$ \cite{37,41}.
Polarization operator $\Pi\left(\omega,\mathbf k\right)$ is
\begin{gather}\label{11}
	\Pi\left(\omega,\mathbf k\right)=i\int G\left(\varepsilon_1,\mathbf p_1\right)
	G\left(\varepsilon_2,\mathbf p_2\right)\frac{\df \varepsilon\df{\mathbf p}}{\left(2\pi\right)^3} \com
\end{gather}
where $G\left(\varepsilon,\mathbf p\right)$ is Green function of electron,
$\mathbf p_{1,2}=\mathbf p\pm\mathbf k/2$,
$\varepsilon_{1,2}=\varepsilon\pm\omega/2$.
After substitution of causal Green function into \eqref{11} we obtain in long-wavelength approximation
\begin{gather}\label{12}
	\Pi\left(\omega,\mathbf k\right)=\frac{k^2}{4\pi\omega^2}\sum\limits_{i=\pm}\left|v\left(p_i\right)\right|p_i \com
\end{gather}
where
$v=\partial_p\varepsilon\left(p\right)$,
$p_i$ is the root of the equation
$\varepsilon\left(p\right)=\varepsilon_\text F$
($i=+,\mspace{3mu}-$, in which connection $p_+$ lies in the region of positive effective mass and $p_-$ lies in the region of negative effective mass, \hyperref[fig1]{Fig. 1})
$\varepsilon\left(p\right)$ is electron spectrum \eqref{9} with upper signs.
The next formula follows from the Eqs. \eqref{10} and \eqref{12}
\begin{gather}\label{13}
	\omega_0=e\sqrt{\frac{\varepsilon_\perp}{2v_\text F}\sum\limits_{i=\pm}\left|v\left(p_i\right)\right|p_i} \pnt
\end{gather}
Using Eq. \eqref{9} we obtain for the electron velocity
\begin{gather}\label{14}
	v=\frac{v_\text F q}{\varepsilon}\left(1-\frac{1+u^2}{\sqrt{1+4\left(1+u^2\right)q^2}}\right) \com
\end{gather}
and for the parameter $\omega_0$
\begin{multline}\label{15}
	\omega_0\left(u\right)=\frac{e\varepsilon_\perp}{\sqrt{2v_\text F}}\mspace{5mu}
	\Theta\left(2\mu-g\right)
	\times \\
	\sqrt{\Big(v\left(u,q_+\right)q_+-v\left(u,q_-\right)q_-\Theta\left(u-2\mu\right)\Big)} \pnt\mspace{30mu}
\end{multline}
Here we define
$q=v_\text F p/\varepsilon_\perp$,
$u=V_0/\varepsilon_\perp$,
$\mu=\varepsilon_\text F/\varepsilon_\perp$,
$g=\Delta_\text g/\varepsilon_\perp$,
$\Delta_\text g$ is determined by the formula \eqref{16},
$\Theta\left(\xi\right)$ is unit step function,
$$
q_\pm=\frac{1}{2}\sqrt{4\mu^2+u^2\pm 2h} \com
$$
$$
h=\sqrt{4\mu^2\left(1+u^2\right)-u^2} \pnt
$$

\begin{figure}[t]
	\centering
	\includegraphics[width=\linewidth]{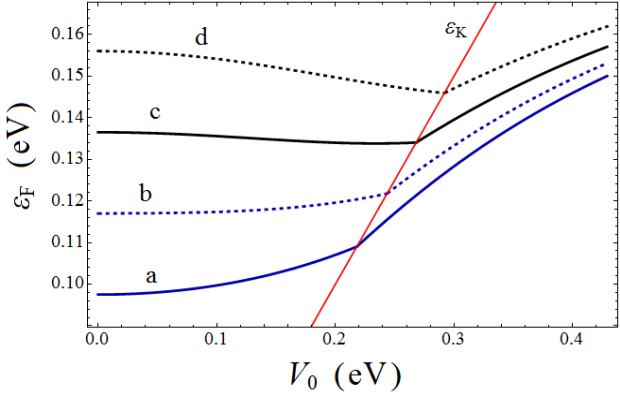}
	\caption{Position of Fermi level $\varepsilon_\text F$ vs bias $V_0$;
		(a) -- $n_0=1.9\cdot10^{12}$ cm$^{-2}$; (b) -- $n_0=2.4\cdot10^{12}$ cm$^{-2}$; (c) -- $n_0=2.9\cdot10^{12}$ cm$^{-2}$; (d) -- $n_0=3.5\cdot10^{12}$ cm$^{-2}$.
		Red line is energy corresponding to K point.}
	\label{fig2}
\end{figure}

We consider doped bigraphene at zero temperature and the surface concentration $n_0$ of free carriers is suggested to be fixed.
To find out the position of Fermi level $\varepsilon_\text F$ we use the standard method of statistical physics which leads to the equation
\begin{gather}\label{1}
p_+^2\left(\varepsilon_\text F\right)-p_-^2\left(\varepsilon_\text F\right)=2\pi n_0 \pnt
\end{gather}
Here we have taken into account that Fermi surface of biased bigraphene is ring.
The result is
\begin{multline}\label{3}
4\mu^2+u^2+2\sqrt{4\mu^2\left(1+u^2\right)-u^2} \\
-\left(4\mu^2+u^2+2\sqrt{4\mu^2\left(1+u^2\right)-u^2}\right)\Theta\left(u-2\mu\right)=A \com
\end{multline}
where $A=8\pi v_\text F^2n_0/\varepsilon_\perp^2$.
Thus according to Eq. \eqref{3} if $n_0$ has fixed value then the position of Fermi level depends on the bias voltage.
The explicit dependence has the form
\begin{multline}\label{4}
\mu\left(u\right)=\frac{1}{2}\sqrt{2+A+u^2-2\sqrt{1+A\left(1+u^2\right)}} \\
\times\Theta\left(\sqrt{A}-2u\right)
+\frac{1}{8}\sqrt{\dfrac{A^2+16u^2}{1+u^2}}\Theta\left(2u-\sqrt{A}\right) \pnt
\end{multline}

\begin{figure}[t]
	\centering
	\includegraphics[width=\linewidth]{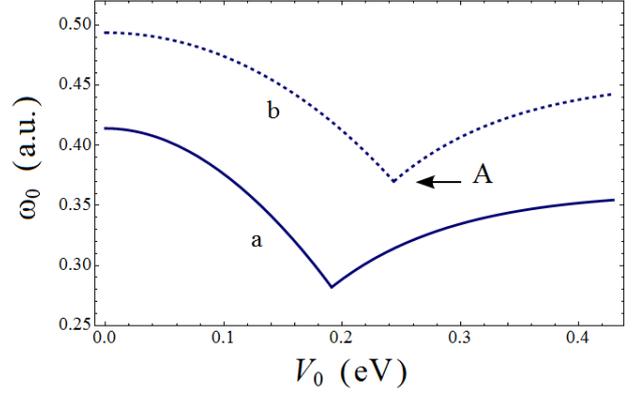}
	\caption{Parameter $\omega_0$ vs bias $V_0$;
		(a) -- $n_0=1.5\cdot10^{12}$ cm$^{-2}$; (b) -- $n_0=2.4\cdot10^{12}$ cm$^{-2}$.
		Point marked with A is breaking point.}
	\label{fig3}
\end{figure}

The position of Fermi level vs the bias is shown in \hyperref[fig2]{Fig. 2}.
We note that values of free carriers concentration $n_0$ used here for numerical analysis are quite comparable with those used in well known experiments previously \cite{18,33}.
To find out the function $\omega_0\left(u\right)$ the Fermi level dependence on the bias (Eq. \eqref{4}) should be taken into account in Eq. \eqref{15}.
Finally, the dependence of the parameter $\omega_0$ on the bias is shown in \hyperref[fig3]{Fig. 3} for different values of free carriers concentration.
It is seen from \hyperref[fig3]{Fig. 3} that the $\omega_0$, and as a consequence, the curvature of the dispersion curve for plasmon $\omega_{\boldsymbol\kappa}$ can be regulated by changing of the bias voltage.
The possibility of this effect had been predicted in Ref. \cite{32} for nondegenerate electron gas.
However the results obtained in Ref. \cite{32} are valid for $u\ll 1$.

Besides, in \hyperref[fig3]{Fig. 3} one can see the possibility of breaking point for the function $\omega_0\left(u\right)$.
In \hyperref[fig3]{Fig. 3} this point is marked with A.
To observe such effect the next requirements should be performed:
(1) the electron gas should be degenerate;
(2) at the initial (minimum) value of the bias voltage the Fermi level should exceed the electron energy in the K point $\varepsilon_\text K$.
In Ref. \cite{32} the electron gas was nondegenerate and Fermi level lied inside the energy gap.
So the appearance of breaking point for the function $\omega_0\left(u\right)$ can't be predicted within the theory \cite{32}.
Indeed, in Ref. \cite{32} the plasma frequency at fixed wave number decreased monotonously with parameter $u$.

The cause of appearance of the breaking point A is as follows.
Let the Fermi level exceeds the electron energy in the K point at the initial value of the bias voltage.
\hfill The \hfill energy \hfill gap \hfill increases \hfill with \hfill the \hfill bias \hfill $u$.
\onecolumn
\begin{multicols}{2}
\noindent It leads to the fact that the region of negative effective mass formed in the vicinity of the K point shifts upward in energy closer to the Fermi level (\hyperref[fig1]{Fig. 1}).
As long as this region remains entirely below the Fermi level electrons from it do not participate in the formation of plasma oscillations.
The contribution to plasma oscillations is provided only by electrons from the region of positive effective mass.
As soon as the Fermi level begins to cross the region of negative effective mass the appearance of additional terms has appeared in the Eq. \eqref{12}.
This leads to a certain increase in the plasma frequency for a fixed value of wave number.

According to Eq. \eqref{4} the breaking point has been reached when the bias voltage becomes equal to
$u_\text A=\sqrt{A}/2$.
With the help of dimensional variables we have $V_{0\text A}=v_\text F\sqrt{2\pi n_0}$.
Thus the bias voltage $u_\text A$ can be controlled by doping.
Namely, the value $u_\text A$ increases with free carriers concentration as $\sqrt{n_0}$.
\\

\textbf{V. Discussion}\\

Above the bias voltage effect on the dispersion curve for plasmon in bigraphene has been investigated.
Note that we have considered the plasmons in the biased bigraphene unlike Refs. \cite{36,37,38,40}.
Also in contrast to \cite{32} we have used the more general electron spectrum which is also valid in the case when $V_0\sim\varepsilon_\perp$.
The curvature of the dispersion curve for plasmon at $k=0$ and the plasmon group velocity have been shown to be regulated by changing of bias voltage.

Dependence of the parameter $\omega_0$ on the bias voltage has been shown above to have nonmonotonous character in opposite to \cite{32}.
The presence of the region of negative effective mass in the vicinity of the K point is the cause for existence of breaking point of the function $\omega_0\left(u\right)$ (point A in \hyperref[fig3]{Fig. 3}).
Indeed, the energy gap increases with the bias $V_0$.
It leads to the fact that the region of negative effective mass formed in the vicinity of K point shifts upward in energy closer to the Fermi level (\hyperref[fig1]{Fig. 1}).
As long as this region remains entirely below the Fermi level electrons from it do not participate in the formation of plasma oscillations.
The contribution to the plasma oscillations is provided only by the electrons from the region of positive effective mass (electrons with quasimomentums $\mathbf p_+$).
As soon as the Fermi level begins to cross the region of negative effective mass (i.e. $\mathbf p_-$ becomes nonzero and electrons with quasimomentums $\mathbf p_-$ contribute to the plasma oscillations also) the additional terms in the Eq. \eqref{12} has appeared.
This leads to an increase in the plasma frequency for a fixed value of wave number.
The contribution of the electrons with pulses $\mathbf p_-$ becomes possible at the finite value of the bias $u_\text A$ which corresponds to the breaking point for the function $\omega_0\left(u\right)$.
The value of the parameter $u_\text A$ increases with free carriers concentration as $\sqrt{n_0}$.

To observe the peculiarity of plasma oscillations described above the Fermi level should exceed the electron energy in the K point at initial value of bias voltage.
Note that in Ref. \cite{32} the electron gas was nondegenerate and Fermi level lied inside the energy gap.
So the appearance of breaking point of the function $\omega_0\left(u\right)$ can't be predicted within the theory \cite{32}.

As mentioned above high value of plasmon velocity in comparison with electron drift velocity is the reason of possible use of plasmon in ultrafast nanoelectronics in the future.
On the other hand the use of bigraphene as the working medium more preferable than the use of single graphene due to the possibility of gap opening in energy structure of biased bigraphene.
However bias increasing can lead to decreasing of plasmon group velocity \cite{32}.
As it has been shown above in doped bigraphene there is the value of the bias $u_\text A$ at which the decrease of the parameter $\omega_0$, and as a consequence, the decrease of group velocity are replaced by their increase.
Such behaviour of group velocity has been explained above by contribution of electrons with negative effective mass to the plasma oscillation.
This contribution is manifested only at a certain value of the bias voltage.

At the end we note that for all carriers concentrations $n_0$ and all values of the bias $V_0$ used above
$\varepsilon_\text F<\varepsilon_\perp$.
Parameter $\varepsilon_\perp$ is equal to minimum for upper branch of dispersion law corresponding to the conduction band.
Thus the absence of accounting of this branch is justified.
\\
\\

\textbf{Acknowledgements}\\

This work was supported by the RF Ministry of Education and Science as part of State Order no. 2014/411, project code 3154, and within the State Task, code 3.2797.2017/4.6.\\
\end{multicols}
\begin{center}
	---------------------------------------------------------------------------------------------------------
\end{center}
\vspace{5mm}

\begin{multicols}{2}

\end{multicols}

\end{document}